\documentclass[reprint,aps,superscriptaddress,nofootinbib,preprintnumbers]{revtex4-1}
\pdfoutput=1
\usepackage{graphicx, amsmath, amssymb, amstext}
\usepackage{relsize, float, multirow}
\usepackage{array, epstopdf, hyperref, }
\usepackage{mathtools,pbox}
\usepackage{dcolumn,color}
\usepackage{soul}
\newcommand{\tauks}{\tau \rightarrow K_s \pi \nu_\tau}
\newcommand{\taupks}{\tau^+ \rightarrow K_s \pi^+ \bar{\nu_\tau}}
\newcommand{\taunks}{\tau^- \rightarrow K_s \pi^- \nu_\tau}

\newcommand{\lag}{{\mathcal{L}}}

\newcommand{\hc}{{\text{h.c.}}}
\newcommand{\sMN}{\sigma^{\mu\nu}}
\newcommand{\smn}{\sigma_{\mu\nu}}

\newcommand{\gev}{~\text{GeV}}
\newcommand{\Rns}{R_{\tau,\text{ns}}}
\newcommand{\Rs}{R_{\tau,\text{s}}}

\begin{document}
%------------------------------------------------------------------------------------
%\title{Explanation of $ \tauks $ CP asymmetry with dimension 8 operator : breaking a `no-go theorem' }
%\title{Tending to tensions in $ \tau $ decay with tensors}
\title{Tensors for tending to tensions in $ \tau $ decays}
%\title{Hints of non-Standard tensor interaction from $\tau$ decay anomalies}
%------------------------------------------------------------------------------------------

\author{Amol Dighe}
\email{amol@theory.tifr.res.in}
\author{Subhajit Ghosh}
\email{subhajit@theory.tifr.res.in}
\author{Girish Kumar}
\email{girishk@theory.tifr.res.in}
\affiliation{Department of Theoretical Physics, Tata Institute of Fundamental Research, Mumbai 400005, India} 
\author{Tuhin S.~Roy}
\email{tuhin@theory.tifr.res.in}
\affiliation{Department of Theoretical Physics, Tata Institute of Fundamental Research, Mumbai 400005, India}
%\affiliation{Theory Division T-2, Los Alamos National laboratory, Los Alamos, New Mexico 87545, USA}

%------------------------------------------------------------------------------------------------------

\preprint{TIFR/TH/19-3}

%------------------------------------------------------------------------------------
\begin{abstract}	
We propose a manifestly gauge invariant effective tensor operator that can account for the $ CP $ asymmetry anomaly in $ \tau  $  decays, contrary to the claim made in literature. Additionally, this operator can also help in resolving the long-standing discrepancy in the value of $ V_{us} $ extracted from inclusive $ \tau $ decays. By construction, the operator evades bounds from neutron electric dipole moment, while keeping the extraction of $ V_{us} $ from exclusive $ \tau $  decays unaffected. We explicitly provide a renormalizable model of flavor symmetries that yields the desired effective tensor operator.
\end{abstract}

%\notoc
%----------------------------------------------------------------------------------------------------------

\maketitle	

Decays of $ \tau $ leptons provide a laboratory for precision tests of the Standard Model (SM) of particle physics~\cite{Davier:2005xq,Pich:2013lsa}. Leptonic $ \tau $ decays are crucial for testing lepton flavor universality and searching for lepton flavor violating decays~\cite{Amhis:2016xyh,Tanabashi:2018oca}.
Hadronic  $ \tau  $ decays offer us the opportunity 
to probe strong interactions in a variety of ways. 
Since the mass of $ \tau $ is fairly above the QCD scale, 
  we can use perturbative results derived in the effective field theory framework. These decays have been used to extract fundamental parameters of SM, such as the strong coupling constant $ \alpha_s $~\cite{Braaten:1991qm,Davier:2008sk,Boito:2014sta,Pich:2016bdg}. As $ \tau $ is massive enough to decay to strange mesons, these decays have also been used to determine the strange quark mass and the $ V_{us} $ element of the Cabibbo--Kobayashi--Maskawa (CKM) matrix~\cite{Gamiz:2002nu,Gamiz:2004ar,Baikov:2004tk}.
  
Consistency across various measurements of the above parameters vouches for the success of SM. However further precision measurements of $ \tau $ decays can uncover hints of new physics (NP) beyond SM, through deviations from SM predictions.
In this work, we take  a critical look at two such long-standing anomalies in hadronic $ \tau $ decays:
\begin{itemize}
	\item 

The determination of $ V_{us} $ from inclusive $ \tau $ decays to final states including the strange quark. 
The value extracted from inclusive decays~\cite{Amhis:2016xyh},
\begin{equation}
|V_{us}|^{\tau,\text{in}} = 0.2186 \pm 0.0021 \; ,
\end{equation}
shows a striking $ 3.1 \sigma$ deviation  from the CKM unitarity prediction,
\begin{equation}
|V_{us}|^\text{uni}=0.22582 \pm 0.00089\; .
\end{equation}
\item 
The CP asymmetry in $ \tauks $ 
%The value of this asymmetry has been calculated 
is calculated in SM to be~\cite{Bigi:2005ts,Grossman:2011zk}
\begin{equation}
A_{CP}^{\tau,\,{ \rm SM}} = (0.33 \pm 0.01)\% \; .
\end{equation}
However the asymmetry measured by the BaBar collaboration~\cite{BABAR:2011aa},
\begin{equation}
A_{CP}^{\tau,\, \text{exp}} = (-0.36 \pm 0.23 \pm 0.11)\% \;,
\end{equation}
 is $ 2.8 \sigma $ below the SM prediction.
\end{itemize}
 
Various attempts have been made in the literature to reconcile the $ V_{us} $ anomaly within SM. Proposals range from inadequate theoretical understanding of $ \tau $ spectral functions~\cite{Boyle:2018ilm,Hudspith:2017vew} to possible mis-measurements of some exclusive $ \tau $ decay channels~\cite{Antonelli:2013usa}. (The measurement of inclusive $ \tau \to s $ decays is through the summation over all exclusive decay modes with an odd numbers of kaons~\cite{Amhis:2016xyh} in the final state.) It is, however, not possible to account for the discrepancy in $ CP $ asymmetry within SM~\cite{Devi:2013gya,Dhargyal:2016kwp,Cirigliano:2017tqn}. It was pointed out in Ref.~\cite{Devi:2013gya} that scalar NP operators cannot generate new $ CP $ asymmetry while tensors can. However Ref.~\cite{Cirigliano:2017tqn} claimed that a demand of gauge invariance brings the tensor operator in conflict with the measurement of neutron electric dipole moment (EDM). This claim has recently been repeated in Ref.~\cite{Rendon:2019awg}.

%The interference of SM and NP  amplitude in principal can yield nonzero direct $ CP $ violation  provided both weak and strong phases survives in the interaction term. In Ref~\cite{Devi:2013gya}, it was first pointed out that while the scalar NP cannot generate $ CP $ asymmetry due to no strong phase, the tensor NP interaction can indeed provide new source of $ CP $ asymmetry. However, as argued in Ref~\cite{Cirigliano:2017tqn}, this NP explanation is strongly in conflict  with data on neutron .... 

In this Letter, we address both the above $ \tau $ anomalies with a single source of NP in the effective field theory (EFT), and provide an explicit renormalizable model. 
The essential requirements we impose are
% proposal will have to contend with the following two requirements : 
$ (i) $ no effect on $ V_{us} $ extraction from exclusive $ \tau $ decays, which is consistent with SM within $ \sim1\sigma $,
%, should remain unaffected even after the introduction of NP, 
$ (ii) $ manifest electroweak (EW) gauge invariance.

%These constraints mentioned above give clear guidelines for building models that accommodate the anomalies mentioned above.
 To comply with the the first requirement, the new operator $\mathcal{O}_\text{NP}$ 
should be such that
\begin{equation}
\langle K \nu_\tau \left| \mathcal{O}_\text{NP} \right|\tau\rangle  = 
\langle K | \mathcal{O}_\text{NP}^\text{had} |0 \rangle ~ \langle \nu_\tau | \mathcal{O}_\text{NP}^\text{lep} |\tau\rangle 
 =  0\;.
\end{equation}
This condition is satisfied automatically if $\mathcal{O}_\text{NP}^\text{had}$, the hadronic part of the NP operator, has a tensorial structure. This is because $\langle K|\bar{q} \ \smn \gamma_5 \ q' |0\rangle = 0$
for any flavors $q,q'$, due to the antisymmetry of the Lorentz structure of the operator.
Note that the symmetry argument would not be valid if the outgoing state were to involve more than one hadron. Therefore, by choosing the flavor structure of  $\mathcal{O}_\text{NP}^\text{had}$ appropriately, one can contribute to inclusive $ \tau $ decays. Thus a suitable choice of $ \mathcal{O}_\text{NP} $ that accounts for the $ V_{us} $ discrepancy would be the tensor operator $\left(\bar{s} \smn u \right) \left(\ \bar{\nu}_{\tau} \sMN \tau \right)$.
%$\tau \rightarrow K + X +\nu_\tau$, where $X$ represents at least one extra hadron. This, in turn, guarantees a non-trivial contribution to inclusive $\tau$ decays to strange mesons. 

%Putting everything together, we conclude that a suitable choice for $\mathcal{O}_\text{NP}$, that does not affect $\tau \rightarrow K^- \nu_\tau$ yet contributes to  $\tau \rightarrow K + X +\nu_\tau$, is of the form $\left(\bar{s} \smn u \right) \left(\ \bar{\nu}_{\tau} \sMN \tau \right)$. 

Remarkably, the same operator can also contribute to the $ CP $ asymmetry in $ \tauks $~\cite{Devi:2013gya}. Therefore, it may be possible to reconcile both these anomalies by a unified explanation. On the other hand, according to Ref.~\cite{Cirigliano:2017tqn}, a demand of EW gauge invariance necessarily gives rise to an additional operator $\left(\bar{u}_L \smn u_R \right) \left(\tau_L \sMN \tau_R \right)$, 
%with similar strength but involving charged leptons instead of neutrinos, 
which is strongly constrained from neutron EDM measurements. Consequently, there appears to be a ``no-go" for the explanation of the $ CP $ anomaly. 

%%In this Letter, we argue that the no go condition mentioned above can be evaded in two ways. One is to have a flavor model such that the CKM rotation originates entirely from the $ d$-type sector. Another method
%, which does not rely upon specific flavor rotation 
%%which we focus on in this paper, is imposing gauge invariance in alternative way.
% would be to have an
%%We propose a concrete example where we append to SM a \emph{single effective tensor operator}
%reasonings as sketched by authors in Ref~\cite{Cirigliano:2017tqn} only pertains to a specific way of imposing gauge invariance. 
%We demonstrate that there exists at least one class of manifestly gauge invariant operators that can accommodate both anomalies in $\tau$ decays and yet is safe from neutron EDM constraints. As a concrete example, we append to SM a \emph{single effective tensor operator}

In this Letter, we argue that the no-go theorem sketched in Ref.~\cite{Cirigliano:2017tqn} only pertains to a specific way of imposing gauge invariance. 
We demonstrate that there exists at least one class of manifestly gauge invariant operators that can accommodate both anomalies in $\tau$ decays and yet is safe from neutron EDM constraints. 
As a concrete example, we propose a renormalizable flavor model (detailed later in this article) which appends to SM the following \emph{gauge invariant effective tensor operator} after integrating out all beyond-SM degrees of freedom:

%As a concrete example, we append to SM a \emph{single gauge invariant effective tensor operator}

\begin{equation}
\lag \supset {\mathcal{K} \over \Lambda^4}\left[(\bar{\ell}_{3}H^\dagger) \smn \tau_{R}\right]\left[(\bar{q}_{2}H) \sMN u_{R}\right] + \hc  \; , 
\label{eqn:EWoperator}
\end{equation}
where $\Lambda$ represents the scale of the operator, defined such that $\left| \mathcal{K} \right| = 1$. 
%\st{We also propose a renormalizable flavor model, (given later) in which,} 
In our model, after electroweak symmetry breaking (EWSB), Eq.~\eqref{eqn:EWoperator} yields a single operator where all the fields are written in the mass basis:
%, writing the Higgs doublet $H$ in the unitary gauge, 
%\st{the above operator reduces to the familiar form }
\begin{equation}\label{eqn:eff}
\lag \supset -\frac{4G_F}{\sqrt{2}} V_{us}C_T\left[(\bar{s}_L~\smn u_R )(\bar{\nu}_{\tau L}\sMN \tau_R)  \right] \;.
\end{equation}
%\st{where all the fields are written in mass basis.} 
We have used
\begin{equation}
C_T\equiv -{\mathcal{  K} \over 2\sqrt{2}V_{us}} {v^2 \over G_F \Lambda^4} = -{\mathcal{K} \over 2 V_{us}}\left( v \over \Lambda \right)^4 \; ,
\end{equation}
with $v = 246\gev$ is the EWSB scale, and $ G_F $ is the Fermi constant.
(Note that the factor of $ V_{us} $ is used in the normalization of $ C_T $, only to simplify further calculations.)
 This corresponds to a 4-fermion interaction with effective coupling strength $  \mathcal{  K} v^2 / 2 \Lambda^4 $. The NP coupling via the Higgs doublet in Eq.~\eqref{eqn:EWoperator} ensures that 
 %we generate only the operator in Eq.~\eqref{eqn:eff} from the gauge invariant Lagrangian, thereby avoiding 
 the dangerous operator that would have contributed to the neutron EDM is not generated.

Before going to the details of the renormalizable model, we now quantitatively explore our solutions to the above two $ \tau $ anomalies, including constraints from other relevant measurements.
% in details and also looked into the constraints coming from measurements of branching ratio to some prominent exclusive decay channels.
In SM, $ \tau $ decays to strange and non-strange hadrons proceed via the standard 4-fermi interaction,
\begin{multline}\label{eq:SM}
\mathcal{L}\supset - {4G_F \over \sqrt{2}}\left[V_{us}(\bar{s}_L\gamma^\mu u_L )(\bar{\nu}_{\tau L}\gamma_\mu \tau_L)\right.\\
\left.+ V_{ud}(\bar{d}_L\gamma^\mu u_L )(\bar{\nu}_{\tau L}\gamma_\mu \tau_L)\right]\;.
\end{multline}
Let us first explore the extraction of $ V_{us} $ from inclusive decay channels.
In this extraction, strange and non-strange spectral functions of $ \tau $ are used along with very precisely determined value of $ V_{ud} $. 
The relevant quantity is~\cite{Braaten:1991qm}
\begin{equation}\label{in1}
\delta R_\tau \equiv {\Rns\over |V_{ud}|^2}-{\Rs \over |V_{us}|^2} \;,
\end{equation}
where $\Rns$ and  $ \Rs $ are the partial widths defined as
\begin{equation}\label{eq:rtsns}
R_{\tau,\text{ns}(\text{s})} \equiv {\Gamma[\tau \rightarrow \nu_\tau+X_{\text{ns}(\text{s})}] \over \Gamma[\tau \rightarrow e \nu_\tau  \bar{\nu}_e]} \;.
\end{equation}
Here $ X_{\text{ns}(\text{s})} $ represents the final state with even (odd) strangeness.
The quantity $ \delta R_\tau $,
which vanishes in the perturbative and massless quark limit,
encodes the corrections due to finite quark masses and non-perturbative effects. Using QCD sum rules and operator product expansion (OPE) \cite{Shifman:1978bx,Shifman:1978by,Craigie:1981jx}, this factor is theoretically calculated in SM to be~\cite{Gamiz:2004ar,Gamiz:2006xx}
\begin{equation}
	\delta R_\tau^\text{SM} = 0.242 \pm 0.032 \;.
\end{equation}
Using this theoretical input and experimentally determined values of $ \Rs $, $ \Rns $, and $ V_{ud} $ \cite{Amhis:2016xyh}, the CKM element $ V_{us} $ is extracted using Eq~\eqref{in1}, and is found to be nearly $ 3.1 \, \sigma $ \cite{Amhis:2016xyh} below the CKM unitarity prediction. 

Now we calculate the effect of our NP on this extraction. We expect that the tensor operator will modify $ \Rs $, thereby changing the theoretical value of $ \delta R_\tau $. 
The effect of this operator on the inclusive mode can be calculated using QCD sum rules and OPE \cite{Craigie:1981jx,Cata:2007ns}. 
We parametrize the corrections due to NP as
\begin{equation}\label{eq:drtNP}
\delta R_\tau^\text{NP}(C_T) = \delta R_\tau -  \delta R^\text{SM}_\tau \;,
\end{equation}
and find
\begin{equation}
\delta R_\tau^\text{NP}(C_T) \approx  -288\pi^2 \text{Re}(C_T){\langle 0|\bar{q}q|0\rangle \over m_\tau^3} - 18|C_T|^2 \;,
\label{eq:Rnp}
\end{equation}
where $ \bar{q}q \equiv (\bar{u}u + \bar{s}s) / 2$. 
In deriving the above, we have ignored higher order correction to OPE due to quark masses, gluon condensates, higher order corrections in $ \alpha_s $, and duality violation \cite{Chibisov:1996wf,Shifman:2000jv,Cata:2005zj} effects. 

The first term in Eq.~\eqref{eq:Rnp} arises from the interference of the SM operator in Eq.~\eqref{eq:SM} with the tensor operator that we introduced in Eq.~\eqref{eqn:eff}, and therefore is linear in $ C_T $. The second term, quadratic in $ C_T $, comes from the tensor-tensor correlator. We briefly sketch the derivation of Eq.~\eqref{eq:Rnp}, and specially the estimation of the tensor-tensor correlator, in Appendix~\ref{sec:app1}.

%\section{CP asymmetry in $\tauks  $}

Let us turn towards the other anomaly, namely, the $ CP $ asymmetry  in $ \tauks $ decays:
\begin{equation}
A^{\tau}_{CP} = {\Gamma(\taupks)-\Gamma(\taunks) \over \Gamma(\taupks)+\Gamma(\taunks)} \;.
\end{equation}
In SM, the source of this asymmetry is indirect $ CP $ violation due to $ K-\overline{K} $ mixing in the final state \cite{Bigi:2005ts}. After the initial inconclusive null results from CLEO~\cite{Bonvicini:2001xz} and Belle~\cite{Bischofberger:2011pw}, the BaBar collaboration has measured this $ CP $ asymmetry at the PEP-II asymmetric-energy $ e^+$-$e^- $ collider at SLAC~\cite{BABAR:2011aa}.

The events in the signal channel $ \tau \rightarrow \pi K^0_S\nu_\tau $(C1) at BaBar also receive significant contamination from the two background channels, $ \tau \rightarrow K K^0_S\nu_\tau $(C2) and $ \tau \rightarrow \pi K^0\bar{K}^0\nu_\tau $(C3).
Therefore, the actual asymmetry measured is~\cite{BABAR:2011aa}
\begin{equation}
\mathcal{A}={f_1 A_1 + f_2 A_2 + f_3 A_3 \over f_1 + f_2 +f_3} \;.
\end{equation}
Here $ A_1 $, $ A_2 $, $ A_3 $ are $ CP $ asymmetries in the channels C1, C2, C3,
respectively, while $ f_1, f_2, f_3 $ are the corresponding sample fractions.
Note that the above $ CP $ asymmetries measured in the experiment are also affected by the $ K_S \rightarrow \pi^+ \pi^- $ decay time dependence of the event selection efficiency~\cite{Grossman:2011zk}. As argued by BaBar, this results in a multiplicative correction factor such that~\cite{BABAR:2011aa}
\begin{equation}\label{key}
A_1 = (1.08 \pm 0.01) A^{\tau}_{CP} \;.
\end{equation}
In SM, $ A_1=-A_2 $, and $ A_3=0 $. Using these relations and the measurement~$ \mathcal{A}^\text{exp}  = (-0.27 \pm 0.18 \pm 0.08)\%$, BaBar extracted $ A_1^\text{exp}=  ( - 0.36 \pm 0.23 \pm  0.11)\% $. The corresponding SM prediction is $A_1^\text{SM} = (+0.36 \pm 0.01) \% $. This is the $ 2.8 \,\sigma $ discrepancy in the $ CP $ asymmetry in $ \tau $ decays.

While introducing NP, a very important point needs to be considered which has been overlooked in previous analyses~\cite{Devi:2013gya,Dhargyal:2016kwp,Cirigliano:2017tqn}. The extraction of $ A_1 $ from $ \mathcal{A} $ as done by BaBar, while valid in SM, fails in the presence of NP. This is because in general, $ A_1 \ne -A_2 $. Therefore the comparison of predictions in the presence of NP needs to be done with the measured quantity $ \mathcal{A} $, and not with $ A_1 $. 
The NP operator as introduced by us in Eq.~\eqref{eqn:eff} affects only the channel C1. Therefore
\begin{equation}\label{eq:As}
A_1 = A_1^\text{SM} + A_1^\text{NP}\;, ~  A_2 = A_2^\text{SM}\;, ~ A_3 = A_3^\text{SM} = 0\;.
\end{equation}
While $ A_1^\text{SM} $ is positive, the direct CP violation arising from interference of the $ V$--$A $ operator in Eq.~\eqref{eq:SM} and the tensor operator in Eq.~\eqref{eqn:eff} can lead to a negative %contribution  the direct $ CP $ asymmetry 
$ A_1^\text{NP}  $. This would move the value of $ A_1 $, and hence the value of $ \mathcal{A} $, closer to the measurement.

%Taking $ f_1,f_2,f_3 $ as given in ~\cite{BABAR:2011aa}, $ \mathcal{A}^\text{SM} = xx.xx \pm yy.yy$ which is $ 2.8\,\sigma $ higher than the measured value.
%
%The NP operator as introduced by us in Eq.~\eqref{eqn:eff} affects only the channel C1. Therefore
%\begin{equation}\label{eq:As}
%A_1 = A_1^\text{SM} + A_1^\text{NP}\;, ~  A_2 = A_2^\text{SM}\;, ~ A_3 = A_3^\text{SM} = 0\;.
%\end{equation}
%If the value of $ A_1^\text{NP} $ is positive, the NP contribution would move the value of $ \mathcal{A} $ closer to the measured value. Note that the indirect $ CP $ violation in the $ K-\bar{K} $ is extremely well measured and consistent with SM. Therefore the NP contribution should be through direct $ CP $ violation, arising from an interference between two operator with different strong and weak phases~[c]. The $ V-A $ operator in Eq.~\eqref{eq:SM} and tensor operator in Eq.~\eqref{eqn:eff} should do the job for us. While the weak phase would be provided by the complex nature of the NP parameter $ C_T $, we will also need the phases $ \delta_+ $(phase of the vector form factor) and $ \delta_T $(phase of the tensor form factor) to be different.

In the presence of both, indirect $ CP $ violation from SM ($ A^{\tau,\text{SM}}_{CP} $) and direct $ CP $ violation from NP ($ A^{\tau,\text{NP}}_{CP} $), the net $ CP $ asymmetry may be written in the form~\cite{Devi:2013gya}
\begin{equation}\label{tacp}
A^{\tau}_{CP} = \frac{A^{\tau,\text{SM}}_{CP} + A^{\tau,\text{NP}}_{CP}}{1+A^{\tau,\text{SM}}_{CP}A^{\tau,\text{NP}}_{CP}}\;.
\end{equation}
The  value of $ A^{\tau,\text{NP}}_{CP}  $ may be expressed in terms of the vector and tensor form factors $ f_+(s) $ and $ B_T(s) $, defined through
\begin{align}
\langle\bar{K}(p_K)\pi(p_\pi)|\bar{s}\gamma^\mu u|0\rangle &= (p_K-p_\pi)^\mu f_+(s) \nonumber\\
&\qquad+ (p_K+p_\pi)^\mu f_-(s)\,,\label{eq:f+-}\\
\langle\bar{K}(p_K)\pi(p_\pi)|\bar{s}\sMN u|0\rangle &= i{p_K^\mu p_\pi^\nu - p_K^\nu p_\pi^\mu \over m_K}B_T(s)\label{eq:bt}\,,
\end{align}
where $ s\equiv (p_\pi + p_K)^2  $. The asymmetry, as may be obtained from the differential decay rate~\cite{Devi:2013gya,Cirigliano:2017tqn} (see Appendix \ref{sec:app2}), is
\begin{equation}\label{acpmag}
A^{\tau,\text{NP}}_{CP} 
=\frac{\text{Im}(C_T)}{\Gamma_\tau \text{BR}(\tauks)}\; \mathcal{I}\;,
\end{equation}
where 
\begin{equation}
\mathcal{I} = \int_{s_{\pi K}}^{m_\tau^2}\kappa(s) |f_+(s) B_T(s)|\sin[\delta_+(s) - \delta_T(s)] ~ds \;.
\end{equation}
Here the denominator $ \Gamma_\tau \text{BR}(\tauks) $ is the average decay width of $ \tau^\pm \rightarrow K_s \pi^\pm {\nu_\tau} $. In the integral $ \mathcal{I} $, we denote $ s_{\pi K} =  (m_\pi + m_K)^2 $, and $ \kappa(s) $ is defined as
\begin{equation}
\kappa(s) = \frac{G_F^2 |V_{us}|^2S_\text{EW}}{256 m_\tau^2 m_K}\frac{(m_\tau^2 -s)^2}{s^2}\lambda^{3/2}(s,m_K^2, m_\pi^2)\;,
\end{equation}
where $ \lambda(a,b,c)=a^2 + b^2 +c^2 -2(ab+bc+ca) $, and~$ S_\text{EW} =1.0194$~\cite{Marciano:1988vm,Braaten:1990ef} is the EW correction factor. Here $ \delta_+(s) $ and $ \delta_T(s) $ are the phases of the form factors $ f_+(s) $ and $ B_T(s) $, respectively.

In the decay $\tau\to K_s\pi \nu_\tau$, the relevant 
form factors are dominated by the $ K^*(892) $ resonance in the elastic limit and $ K^*(1410) $ in the inelastic region. 
Both these resonances can have contributions from $ f_+(s) $ as well as $ B_T(s) $, with $ \delta_+(s) = \delta_T(s) $ in the elastic limit as guaranteed by Watson's final-state theorem~\cite{PhysRev.95.228}. Therefore only the inelastic region contributes to $ \mathcal{I} $, and hence to the $ CP $ asymmetry.

The value of $ \mathcal{I} $ is highly uncertain, mainly due to lack of information on $ B_T(s) $. Assuming only SM, the value of $ f_+(s) $ is obtained by fitting to a superposition of Breit-Wigner (BW) functions,~\cite{Epifanov:2007rf} or by using the dispersive parametrization~\cite{Boito:2008fq,Boito:2010me,Antonelli:2013usa}. In the context of NP, Ref.~\cite{Cirigliano:2017tqn} uses simplified ``Omn\`es function" forms \cite{Omnes:1958hv} for the magnitudes of $ f_+(s) $ and $ B_T(s) $, with the phase difference $ \delta_+(s)-\delta_T(s) $ proportional to the BW phase of $K^\ast (1410)$. The constraint $|A^{\tau,NP}_{CP}| \lesssim 0.03 \,|\text{Im} ~C_T|$, obtained therein, is crucially dependent on the value of $ \delta_+(s) $ used, and on the assumed relation $\delta_T(s) = -\delta_+(s)$ in the inelastic region, which has no strong theoretical motivation.
Here we employ a more general treatment, with the same Omn\`es forms for $ |f_+(s)| $ and $ |B_T(s)| $, but with
\begin{equation}\
\delta_T(s) - \delta_+(s)=  \alpha \times \text{Arg}[\text{BW}(K^\ast (1410))] \;,
\end{equation}
which is automatically consistent with Watson's final-state theorem~\cite{PhysRev.95.228}.
The free real parameter $ \alpha $ acts as a proxy for our ignorance about $ \mathcal{I} $, and hence we can study the dependence of NP constraints for different values of $ \alpha$~(hence for different values of $ \mathcal{I} $). We have taken  $f_+(0) |V_{us}| = 0.2166(5)$~\cite{Antonelli:2009ws}, whereas $  B_T(0) / f_+(0) = 0.676(27)$ \cite{Baum:2011rm} from  lattice calculations.

\begin{figure}[t!]
   	\includegraphics[width=0.93\linewidth]{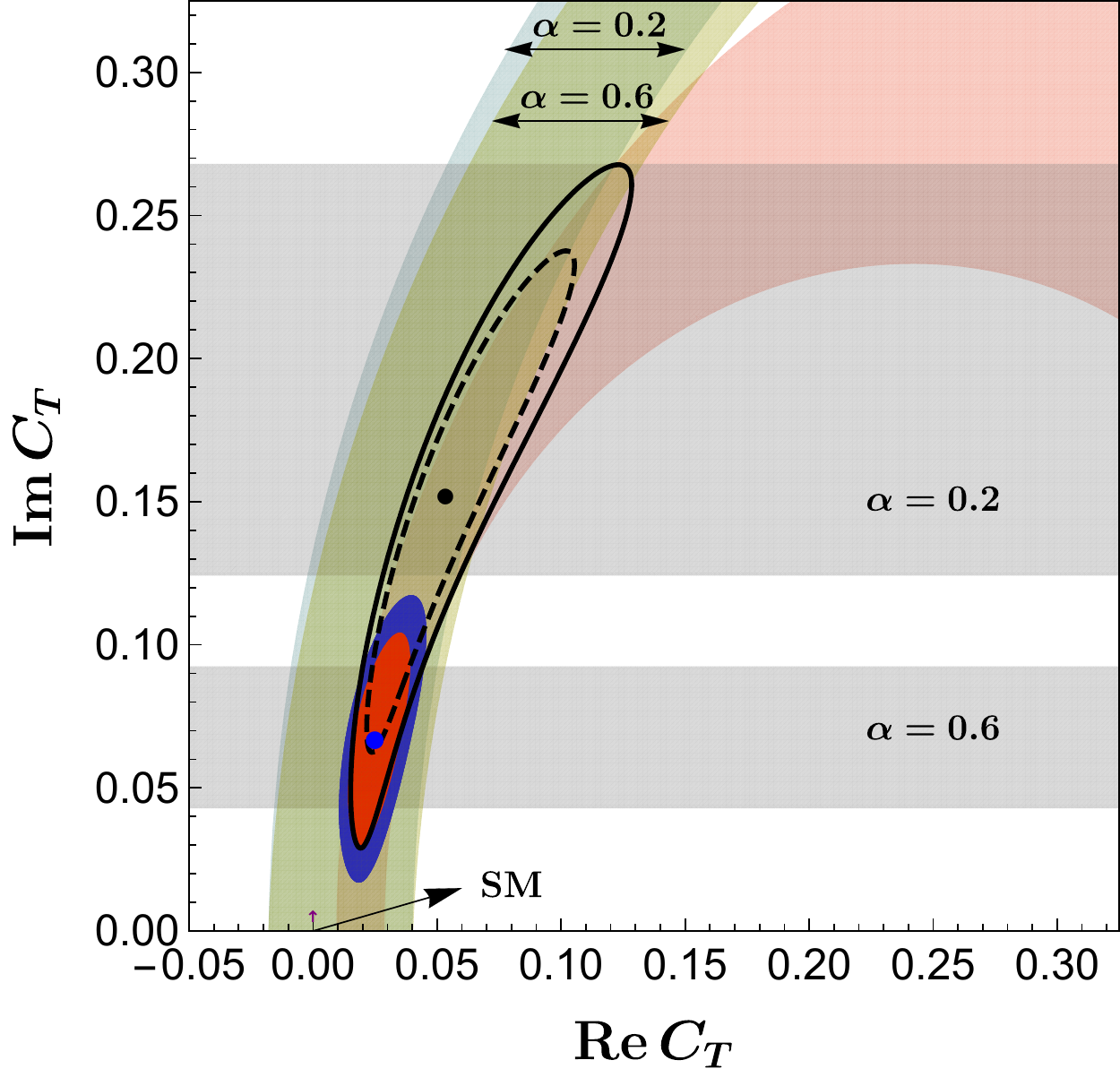}
	\caption{ The $ 1\sigma $ allowed regions from $ A_\text{CP}^\tau $ (grey), BR of $  \tauks$  (light blue, light green) and, the $ V_{us} $ anomaly (pink). The combined $ 1 \sigma $ and $ 90\% $ C.L. regions for $ \alpha = 0.2 $ (unfilled ovals) and $ \alpha = 0.6 $ (filled ovals) are also shown.}
	\label{fig:summary}
\end{figure}

We show our numerical results of the constraints on real and imaginary parts of $C_T$ in Fig.~\ref{fig:summary}.  
The $ 1 \sigma $ allowed regions ($ 1 $ d.o.f, $ \Delta \chi^2 \le 1 $) explaining $ A_\text{CP}^\tau $, for two representative $ \alpha $ values $ (\alpha=0.2,0.6) $, are shown with horizontal grey bands. Note that all our results are invariant under $ \{\alpha\rightarrow - \alpha, ~\text{Im}(C_T)\rightarrow - \text{Im}(C_T)\} $. The value $ \alpha=0.2 $ approximately corresponds to the phase choices in Ref. \cite{Cirigliano:2017tqn}.

The NP introduced will also affect the decay rates of  $\tau \to K_s\pi\nu$ and  $\tau \to K^-\pi^0\nu$. Since the theoretical predictions for these quantities are isospin-related, and hence completely correlated \cite{Antonelli:2013usa}, we only use the branching ratio (BR) measurement for $ \tau \to \bar{K}^0\pi\nu =0.8386\pm 0.0141$~\cite{Amhis:2016xyh}.
The SM value of the BR is taken from~\cite{Antonelli:2013usa}. 
The $1\sigma$ constraints ($ 2 $ d.o.f, $ \Delta \chi^2 \le 2.3 $) in the $C_T$ plane are shown in Fig.~\ref{fig:summary} for $ \alpha=0.2~\text{and }0.6 $, with light blue and light green bands, respectively.

Fig.~\ref{fig:summary} also shows the $1\sigma$ ($ 2 $ d.o.f, $ \Delta \chi^2 \le 2.3 $) allowed regions explaining the $ V_{us} $ anomaly from inclusive $\tau$ decays. The values of quark condensates are taken from Ref.~\cite{McNeile:2012xh} and of other relevant quantities from Refs.~\cite{Amhis:2016xyh,Tanabashi:2018oca}. We have included $10\%$ theoretical errors in the coefficients of the two terms in Eq.~\eqref{eq:Rnp}.

We finally show the combined results from $  A_{CP}^\tau 
$, the BR of $\tau\to K_s\pi\nu$, and the $ V_{us} $ anomaly, for $ \alpha=0.2,0.6 $. Clearly the range of preferred values of NP parameters depends of $ \alpha $. 
For example, for $ \alpha = 0.6  $, the best-fit value of $ \text{Im}(C_T) $  is $ 0.067 $, which is substantially lower than the required value implied in Ref.~\cite{Cirigliano:2017tqn}. Note that this best-fit point has $ |C_T|\approx 0.072 $, which corresponds to $ \Lambda \sim 580\, \text{GeV} $. However, a reliable prediction for $ \mathcal{I} $ is needed to pin down the scale of NP. 

%the value of $ \text{Re}(C_T) $ can be as high as $ 0.12 $ and $ \text{Im}(C_T) $ can be as high as $ 0.25 $ at $ 90\% $ C.L. Since  $ |C_T| $ can be as high as $ 0.3 $, the scale of NP may be as low as $ \Lambda \sim xx \text{GeV} $. 

We now provide a renormalizable model of flavor symmetries that generates the effective operator in Eq.~\eqref{eqn:EWoperator}.
This model preserves the subgroup
%n particular, we  use a subgroup of the full $U(3)^5$ global flavor symmetry of SM, namely:
%
\begin{equation}
\begin{split}
&G =  \left( \prod_{i=1}^3 U(1)_{q_i} \times U(1)_{u_i} \right)\!  \times  U(3)_d 
\\ & \qquad \qquad
\times U(2)_{\ell} \times U(2)_{e} \times U(1)_{\ell_3}  \times U(1)_{e_3} 
\end{split}
\label{eq:symmetry}
\end{equation}
of the $U(3)^5$, which is the global flavor symmetry of SM in the absence of Yukawa couplings. Here we use \emph{left-handed} notation for representing all fermionic fields.
The quark doublets $q_{i}$ have charges $+1$ under $U(1)_{q_i}$, while the quark singlets $u^{c}_{i}$ have charges $-1$ under $U(1)_{u_i}$. The remaining quark singlets $d^{c}_{i}$ form a triplet of $U(3)_d$. The lepton doublets $\ell_i$ and singlets $e^c_i$ from the first two generations are doublets under  $U(2)_{\ell}$ and $ U(2)_{e}$, respectively. The leptons in the third generation, $\ell_3$ and $e^c_3$, carry charges $+1$ and $-1$, under $U(1)_{\ell_3}  $ and $ U(1)_{e_3}$, respectively.

%The pattern of flavor symmetries in Eq.~\eqref{eq:symmetry} is manifest in SM in the limit all Yukawa couplings  of SM (namely $Y_u$, $Y_d$, and $Y_e$) are neglected. 

The additional particles in our model are four Weyl fermions $B, B^c, N, N^c$, and two scalars $\Phi_c, \Phi_y$. The gauge and flavor quantum numbers of these particles are given in Table~\ref{table:1}.
\begin{table}[t!]
	\centering
	\begin{tabular}{c |  c | c  c  c  c|}
		& $\left(SU(3)_C,SU(2)_W\right)_Y  $ & $U(1)_{q_2}$ & $U(1)_{u_1}$  & 
		$U(1)_{\ell_3}$  & $U(1)_{e_3}$ \\
		\hline 	
		$B$ & ~\quad$\left(3,1\right)_{-1/3}  $ & 0 & 1 & 0 & 0 \\
		$B^c$ & ~~$\left(\bar{3},1\right)_{1/3}  $ & -1 & 0 & 0 & 0 \\
		$N$ & $\left(1,1\right)_{0}  $ & 0 & 0 & -1 & 0 \\
		$N^c$ & $\left(1,1\right)_{0}  $ & 0 & 0 & 0 & 1 \\
		$\Phi_c$ & ~~$\left(3,1\right)_{2/3}  $ & 0 & +1 & 0 & -1 \\
		$\Phi_y$ & $\left(1,1\right)_{1}  $ & 0 & 0 & 0 & 0 \\
		\hline
	\end{tabular}
	\caption{Charge assignments of new particles}
	\label{table:1}
\end{table}
%
%The fundamental assumption in our model is that all marginal interactions involving NP particles satisfy the flavor symmetry in Eq.~\eqref{eq:symmetry}. 
The full set of marginal NP interactions are given as 
\begin{equation}
\begin{split}
\lag \ \supset \ &k_1 \: H^\dag q_2 B^c + k_2 \: H \ell_3 N 
+ k_3 \: \Phi_c u_1^c N^c  
\\
&+  k_4 \: \Phi_c^\dag e_3^c B  
+ k_5 \: \Phi_y  u_1^c B +  k_6 \: \Phi_y^\dag e_3^c N^c  \\
&+ k_7 \: \left|H \Phi_c\right|^2   + k_8 \: \left|H \Phi_y\right|^2 
+ k_9 \: \left|\Phi_c \Phi_y\right|^2\;.
\end{split}
\label{eq:NPmar}
\end{equation} 

Scalar masses preserve flavor symmetries. In order to give masses to NP fermions,
we add two extra spurions $M_B$ and $M_N$, both charged $\left( +1, -1\right)$, under symmetries $\left(U(1)_{q_2}, U(1)_{u_1}\right)$ and   $\left(U(1)_{\ell_3},U(1)_{e_3}\right)$, respectively. 
%These allow two additional relevant operators apart from flavor preserving scalar masses
\begin{equation}
\lag \ \supset \ M_B \: B B^c \ + \ M_N \: N N^c  + \ m_c^2 \: \left| \Phi_c\right|^2
+ \ m_y^2 \: \left| \Phi_y\right|^2 \;.
\label{eq:NPrel}
\end{equation} 

\begin{figure}[b]
	\includegraphics[width=\linewidth]{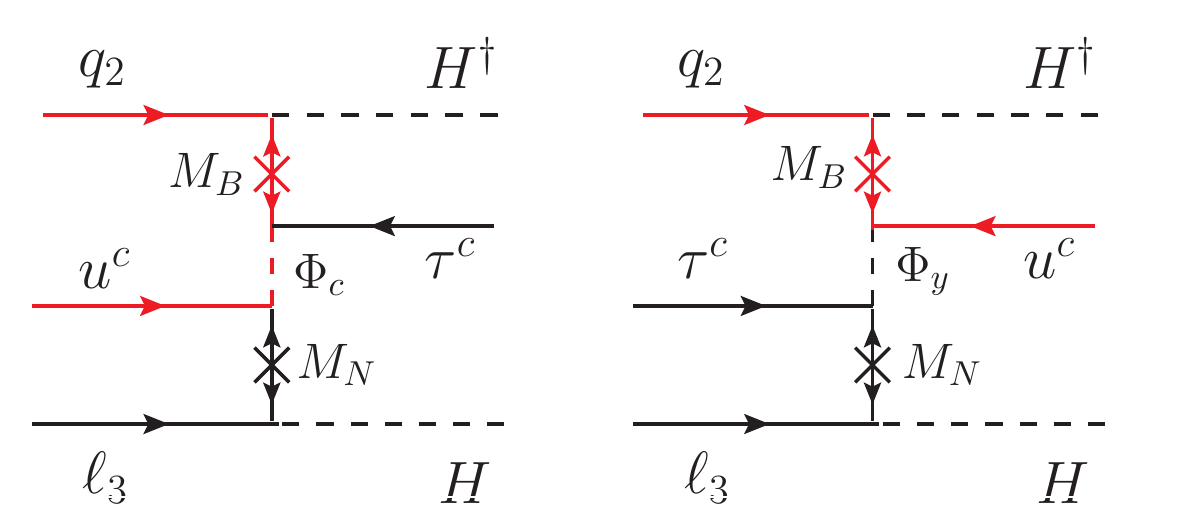}
	\caption{Feynman diagrams that generate NP effective operators responsible for $ \tau $ decays.}
	\label{fig:UVmodel}
\end{figure} 

The SM Yukawa couplings break $ G $  to $ U(1)_B \times \prod_{i=1}^{3} U(1)_{L_i} $, the familiar baryon number and flavor lepton number symmetries. In our scheme, the down-type Yukawa coupling matrix $Y_d$ is such that it can be diagonalized by using only the $ U(3)_d $ symmetry. The up-type Yukawa matrix $ Y_u $ will then have the form $V^\dag_{\text{CKM}} \times \text{diag}\left(y_u, y_c, y_t \right)$. The charged-lepton Yukawa matrix $ Y_e $ is diagonalizable only using the $ U(2)_{\ell} \times U(2)_{e} $ symmetry. This ensures that flavor rotations only affect operators involving left-handed up-type quarks, so that no extra operators are generated along with Eq.~\eqref{eqn:eff}. These considerations would guide the mechanism that generates the flavor structure at a high scale.

%Such Yukawa coupling may be generated through appropriate spurions
%
%, by itself, breaks full quark symmetries to $U(1)_{u}^3 \times U(1)_{B}^3$, and finally to $U(1)_{B}$ along with $Y_u$. Therefore, without loss of generality, we can set 
%$Y_d = \text{diag}\left(y_d, y_s, y_b \right)$ and  $Y_u$ to be $V^\dag_{\text{CKM}} \times \text{diag}\left(y_u, y_c, y_t \right)$. Additionally, we take $Y_e$ to break lepton symmetries to $U(1)_L^3$ and therefore, it can be takes to be $\text{diag}\left(y_e, y_{\mu}, y_{\tau} \right)$.
Given the full set of our symmetries and symmetry-breaking spurions, there are no additional counter-terms.   
%The IR-EFT is derived from Eqs.~(\ref{eq:NPmar}--\ref{eq:NPrel}). 
%At tree level matching one finds the operator given in Eq.~\eqref{eqn:EWoperator}. 
In Fig.~\ref{fig:UVmodel} we show Feynman diagrams where the key effective operators are generated. The left diagram gives rise to $\left(q_{2}H^\dagger e_3^c \right)\left(\ell_{3}H  u_1^c\right)$, which upon Fierz transformation yields the tensor operator in  Eq.~\eqref{eqn:EWoperator} along with its scalar version $\left(q_{2}H^\dagger u_1^c \right)\left(\ell_{3}H  e_3^c\right)$.
After matching we find 
\begin{equation}
\frac{\mathcal{K}}{\Lambda^4} \ = \ - \frac{k_1 k_2 k_3 k_4}{2 M_B M_N m_{c}^2} \;.
\end{equation}   
 The diagram on the right of Fig.~\ref{fig:UVmodel}, on the other hand, gives rise to only the scalar operator $\left(q_{2}H^\dagger u_1^c \right)\left(\ell_{3}H  e_3^c\right)$ with a different effective coupling. Therefore, there exist regions in the parameter space of the EFT  where only the tensor operator dominates, thus realizing our scenario.

%In the parameter space, therefore, there exist patches where the IR-EFT gets characterized by the tensor operator without its scalar counterpart. 

Note that our scenario also generates two additional operators of mass dimension 6, which would give rise to anomalous $ Z\nu_\tau \nu_{\tau}$ couplings. These would constrain the combination $\left(k_2/M_N\right)$ from
invisible $ Z $	width measurements~\cite{Tanabashi:2018oca}. There will also be a series of dimension-8 operators, some breaking lepton flavor universality, which may have important phenomenological consequences relevant for collider experiments like LHC and Belle-II. However, the full analysis of this specific UV model is beyond the scope of this paper and we leave it for future endeavors.

%	, which can give interesting constraints  $\left(k_2/M_N\right)^2$

%
%Apart from the necessary operators (of mass dimension 8), the IR also contains two additional operators of mass dimension 6, which may give rise to anomalous couplings of left handed neutrinos to $Z$, which can give interesting constraints  $\left(k_2/M_N\right)^2$, and a series of dimension-8 operator which have important phenomenological consequences relevant for the LHC.
%However, the full analysis of this specific UV model is beyond the scope of this paper and we leave it for future endevours.  

To summarize, the renormalizable flavor model proposed by us can account for the anomalies in $ V_{us} $ and $ A_{CP}^\tau $ measurements, evading the no-go bound for the latter. Further exploration of these anomalies and the model could unearth hitherto uncharted ways of probing the flavor generation mechanism at a high scale, via low energy flavor data.

\section*{Acknowledgements}

TSR was supported in part by the Early Career Research Award by Science and Engineering Research Board, Dept. of Science and Technology, Govt. of India (grant no. ECR/2015/000196). We also acknowledge the workshop ``Beyond the Standard Model: where do we go from here?'' hosted at the Galileo Galilei Institute for Theoretical Physics where part of the work was completed. 

%the structure of high scale NP through low energy flavor data.

%--------------------- Appendix I -----------------------------%

\appendix

\section{ Correction to $\delta R_\tau $ due to NP}\label{sec:app1}

%The correction to $ \delta R_\tau $ due to NP, as given in Eq~\eqref{eq:drtNP}, is 
%\begin{equation}
%\delta R_\tau^\text{NP}(C_T) = \delta R_\tau -  \delta R^\text{SM}_\tau \;.
%\end{equation}
With the inclusion of the NP tensor operator, $ \Rs $ [see Eq.~\eqref{eq:rtsns}] will receive additional contributions from vector-tensor interference
 %\changer{(why not axial vector- axial tensor?)} \
 and tensor-tensor correlators. The net correction to $ \delta R_\tau $ is 
\begin{multline}\label{drt}
\delta R_\tau^\text{NP}={12\pi^2 \over m_\tau^2}\int_{0}^{m_\tau^2}ds\left(1-{s \over m_\tau^2}\right)
\left[12 ~Re(C_T){\rho_{TV}(q^2) \over m_\tau} \right.\\
\left.- 16|C_T|^2\left(1+{s \over 2m_\tau^2}\right)(\rho^{(Q)}_{TT}(q^2) + \rho^{(R)}_{TT}(q^2)\right]\;,
\end{multline}
with $ s=q^2 $ being the invariant mass of the hadronic states. To derive the above expression we have used the following spectral functions: 
\begin{align}
\rho_{TV}^{\mu\nu\alpha}(q)=\int d\Pi_n(2\pi)^4&\delta^3(q-p_n)\times \nonumber\\
&\sum_n  \langle 0| T^{\mu\nu}|n\rangle \langle n| V^{\alpha\dagger}|0\rangle\;,\\
\rho_{TT}^{\mu\nu,\alpha\beta}(q)=\int d\Pi_n(2\pi)^4&\delta^3(q-p_n) \times \nonumber\\
&\sum_n  \langle 0| T^{\mu\nu}|n\rangle \langle n| T^{\alpha\beta\dagger}|0\rangle\;,
\end{align}
with $  V^{\alpha} \equiv \bar{u}\gamma^\alpha s $ and $ T^{\mu\nu} \equiv  \bar{u}\sMN s $. The spectral functions can be Lorentz decomposed as~\cite{Craigie:1981jx}
\begin{align}
\rho_{TV}^{\mu\nu\alpha}(q) &= i (g^{\mu\alpha}q^\nu - g^{\nu\alpha}q^\mu)\rho_{TV}(q^2)\;,\\
\rho_{TT}^{\mu\nu,\alpha\beta}(q) &= Q^{\mu\nu,\alpha\beta}(q) \rho_{TT}^{(Q)}(q^2)+ R^{\mu\nu,\alpha\beta}(q)\rho_{TT}^{(R)}(q^2)\;,
\end{align}
where
\begin{align}
Q^{\mu\nu,\alpha\beta}(q) =
&(q^\mu q^\beta g^{\nu\alpha}+ q^\nu q^\alpha g^{\mu\beta} -  q^\mu q^\alpha
g^{\nu\beta} - q^\nu q^\beta g^{\mu\alpha})\;,\\
R^{\mu\nu,\alpha\beta}(q) = & Q^{\mu\nu,\alpha\beta}(q) + q^2 (g^{\mu\alpha}g^{\nu\beta} - g^{\mu\beta}g^{\nu\alpha} ) \;.
\end{align}
Using the QCD sum rule framework, the integral in \eqref{drt} can be written as a contour integral on the circle $ {s=m_\tau^2} $, in terms of the associated correlators $ \Pi_{TV} $ and $ \Pi_{TT} $.
\begin{multline}\label{eq:drtint}
\delta R_\tau^\text{NP}={6\pi i \over m_\tau^2}\oint_{s=m_\tau^2}ds\left(1-{s \over m_\tau^2}\right)
\left[12 ~Re(C_T){\Pi_{TV}(q^2) \over m_\tau} \right.\\
\left.- 16|C_T|^2\left(1+{s \over 2m_\tau^2}\right)(\Pi^{(Q)}_{TT}(q^2) + \Pi^{(R)}_{TT}(q^2)\right]\;.
\end{multline} 
To compute the contour integral,
we use the following terms in OPE of the correlators \cite{Craigie:1981jx} in the large-$ q^2 $ limit:
\begin{align}
\Pi_{TV}(q^2)  &\underset{q^2 \rightarrow \infty}{\approx}  {2 \over q^2} \langle0|{1 \over 2}(\bar{u}u + \bar{s}s)|0\rangle\;,\\
\Pi^{(Q)}_{TT}(q^2) = \Pi^{(R)}_{TT}(q^2)  &\underset{q^2 \rightarrow \infty}{\approx} -{N_c \over 24\pi^2}\ln(-q^2) \;,
\end{align}
where $ N_c(=3) $ is the color factor. Evaluating the integral in Eq.~\eqref{eq:drtint}, we obtain Eq.~\eqref{eq:Rnp}:
\begin{align}
 \delta R_\tau^\text{NP}&\approx -288\pi^2 Re(C_T){\langle0|{1 \over 2}(\bar{u}u + \bar{s}s)|0\rangle \over m_\tau^3} - 18|C_T|^2\;.
\end{align}
%---------------------------------------------------------------%

\section{Decay rate of $ \tauks $}\label{sec:app2}

The differential decay rate of $ \tauks $ in the presence of NP is given by~\cite{Cirigliano:2017tqn,Kuhn:1992nz}
\begin{align}\label{eq:brtks}
{d \Gamma \over ds} = G_F^2 |V_{us}|^2S_{EW}\frac{(m_\tau^2 - s)^2 \lambda^{1/2}(s)(m_K^2 - m_\pi^2)^2}{512 m_\tau s^3}\times\nonumber\\
\left[|f_0(s)|^2 + 
\zeta(s)\left(|f_+(s) - T(s)|^2 + \frac{2(m_\tau^2 -s)^2}{9sm_\tau^2}|T(s)|^2\right)\right]
\end{align}
where $ \lambda(s) \equiv \lambda(s,m_K^2, m_\pi^2) $,  $ S_\text{EW}$,  %=1.0194$\cite{Marciano:1988vm,Braaten:1990ef} is the EW correction factor as defined in the main text. The 
and the form factors $ f_+(s),f_-(s) $ and $ B_T(s) $ are defined in Eq.~\eqref{eq:f+-} and \eqref{eq:bt}. The remaining scalar form factor $ f_0(s) $ can be related to the above form factors as
\begin{equation}
f_-(s) = {m_K^2 - m_\pi^2 \over s}\left(f_0(s)-f^+(s)\right)\;.
\end{equation}
The other quantities are defined as
\begin{align}
\zeta(s)&\equiv \frac{(m_\tau^2 + 2s)\lambda(s)}{3 m_\tau^2 (m_K^2 - m_\pi^2)}\;,\\
T(s) &\equiv {3 s m_\tau \over m_K(m_\tau^2 + 2s)}C_T B_T(s)\;.
\end{align}
%Note that the axial part of the operators do not contribute to this process due to parity considerations.

%-----------------------------------------------------------------------------
%\bibliographystyle{unsrtads}
\bibliographystyle{JHEP}
\bibliography{taudecay}

\end{document}